\documentclass[a4paper,12pt]{article}
\usepackage[utf8]{inputenc}
\usepackage[english]{babel}
\usepackage[left=2cm,right=2cm,top=2cm,bottom=2cm]{geometry}

\usepackage{amsmath,amsfonts,amssymb,amsthm}
\usepackage{bm,color}
\usepackage{placeins}
\usepackage{graphicx}
\usepackage{enumerate}
\usepackage{mathtools}
\usepackage{float}
\usepackage[hidelinks]{hyperref}

\usepackage{pdfpages}
\usepackage{lscape}
\usepackage{supertabular}
\usepackage{comment}

\usepackage{tikz}
\tikzstyle{mybox} = [draw=black, very thick, rectangle, rounded corners, inner ysep=5pt, inner xsep=5pt]

\usepackage{caption}
\usepackage{subcaption}

 \usepackage{listings}
\usepackage{textcomp}

\numberwithin{equation}{section}
\title{\textbf{Ballistic electron transport described by a fourth-order Schr\"{o}dinger equation}}
\date{}
\author{Giulia Elena Aliffi\thanks{University of Catania, Department of Mathematics and Computer Science, Catania, Italy  ({\tt giuliaelena.aliffi@phd.unict.it}).}
\and 
Giovanni Nastasi\thanks{University of Enna "Kore", Department of Engineering and Architecture, Enna, Italy ({giovanni.nastasi@unikore.it}).}
\and Vittorio Romano\thanks{University of Catania, Department of Mathematics and Computer Science, Catania, Italy  (Corresponding author: {\tt vittorio.romano@unict.it}).} }
\newcommand{\R}{\mathbb{R}}
\newcommand{\C}{\mathbb{C}}

\newtheorem{proposition}{Proposition}

\newtheorem{remark}{Remark}

\begin{document}

\maketitle

\begin{abstract}
A fourth-order Schr\"{o}dinger equation for the description of charge  transport in semiconductors in the ballistic regime is proposed with the inclusion of non-parabolic effects
in the dispersion relation in order to go beyond the simple effective mass approximation. Similarly to the standard (second order) Schr\"{o}dinger equation, the problem is reduced to a finite spatial domain with appropriate transparent boundary conditions
to simulate charge transport in a quantum coupler \cite{LeKi,AbDeMa,Ab}, where an active region representing an electron device is coupled to leads which take the role of reservoirs. 
Some analytical properties are investigated, and a generalized formula for the current is obtained. Numerical results show the main features of the solutions of the new model. In particular, an effect of interference appears due to a richer wave structure than that arising for the second-order Schr\"{o}dinger equation in the effective mass approximation. 
\end{abstract}


\section{Introduction}
The enhanced miniaturization of modern electron devices makes mandatory to adopt a full quantum description of the charge transport. Among the main approaches for nanoscale devices, in the literature we find the use of the Wigner equation \cite{NeKoSeRiFe,QuDo,Mu,Ja,Ju}, the nonequilibrium Green function \cite{LaDa} and the Schr\"{o}dinger equation \cite{AbDeMa,Ab}. The latter has been employed, for example, for the simulation of resonant tunneling diodes \cite{LeKi,MeJuKo,HeWa,Pi}  in the ballistic regime, giving good characteristic curves, at least from a qualitative point of view. However, the Schr\"{o}dinger equation is almost always employed in the effective mass approximation \cite{HaVa} obtained with a renormalization of the bare electron mass by a factor leading to a reduced mass that encompasses the presence of the periodic potential the charge carriers undergo inside the crystal lattice. 
The main aim of this article is to include non-parabolic effects in the description of electron transport in nanoscale devices in order to go beyond the standard parabolic band approximation within the context of the Schr\"{o}dinger equation. In fact, already in the semiclassical case the simple parabolic band approximation leads to an overestimation of the current, and it is considered as not very accurate \cite{Da,FiLa,Ro}. We expect a similar effect also in a full quantum setting. The complete expression of the dispersion relation in a semiconductor can be obtained by numerical approaches \cite{Ja,FiLa}. However, some analytical approximations  are often adopted with results that improve those based on the simple parabolic band. In this paper, the Kane dispersion relation is considered. Including the complete Kane dispersion relation in the Schr\"{o}dinger equation is a daunting task; here we retain some first order effects beyond the effective mass approximation obtaining a fourth-order Schr\"{o}dinger equation. Moreover, to take into account the additional non-parabolic term in the dispersion relation, we have formulated a generalization of the Transparent Boundary Conditions (TBCs), already devised in \cite{LeKi,AbDeMa,Ab} for the standard (second order) Schr\"{o}dinger equation. 

We highlight that in literature fourth-order Schr\"{o}dinger equations have already been considered in other contexts, e.g., in \cite{KaSha} for  higher-order dispersion and in \cite{BoDroFiWe} for spin current.  Recently some attempts have been also carried on to get a higher-order Schr\"{o}dinger equation  in terms of a real-valued wave function starting from some ideas present in the original paper by Schr\"{o}dinger himself \cite{Schr}; with this spirit in \cite{MaDar} a fourth-order equation for a real "wave function" is proposed, which resembles some analogy with certain equations of elasticity. However, we remark that the underlying idea of these papers is different from the subject of our work because we want to incorporate general dispersion in the canonical Schr\"{o}dinger equation with a standard (complex-valued) wave function as common in solid state physics.

In the present article a one-dimensional case will be investigated. We will be able to reduce the problem of solving the Schr\"{o}dinger equation only in the active area with the new TBCs, obtaining a generalized Sturm-Liouville problem of fourth-order. 
Some analytical properties of the new model are examined. It is proved that the resulting boundary value problem is well posed and an efficient numerical strategy is devised.
Numerical results for a single particle under several kinds of potential - single step, single and  double barrier, double barrier plus a linear potential - show that the new model reveals new features, in particular an effect of interference due to a richer wave structure than that arising for the  Schr\"{o}dinger equation in the effective mass approximation. 

The plan of the paper is as follows: In Sec. 2 the  fourth-order Schr\"{o}dinger equation is obtained by including the effect of non-parabolicity up to the first order. In Sec. 3, the appropriate open boundary conditions are deduced, and in Sec. 4 the well-posedness of the resulting boundary value problem is proved under suitable conditions. Section 5 is devoted to the expression of the current for the  fourth-order Schr\"{o}dinger equation, while in the last section several numerical simulations are presented to show the difference with the standard Schr\"{o}dinger equation in the effective mass approximation.

\section{Fourth-order Sch\"odinger equation}
The general form of the dispersion relation in a semiconductor is obtained by solving the single-electron  Sch\"odinger equation under a periodic potential by employing the Bloch's theorem \cite{Ja,Ju,Lun}.  The complete form can be obtained only numerically, e.g., with the use of pseudo-potential \cite{CheCo,FiLabis}. However, often in the applications analytical approximations are adopted \cite{Da}. Among these, one commonly used is the Kane dispersion relation, which in its isotropic version is given in implicit form as
\begin{equation}
\epsilon (k) (1+\alpha\epsilon( k) )=\frac{\hbar^2 k^2}{2m^*}:=\gamma^2 \ ,
\end{equation}
where $\epsilon(k)$ is the electron energy, $k$  is the modulus of the electron wave-vector and $m^*$ is the effective electron mass; for example, for GaAs  $m^* = 0.067 m_e$ with $m_e$ bare electron mass. The positive parameter $\alpha$ is named non-parabolicity factor, since in the limit $\alpha \to 0^+$ one recovers the standard parabolic band approximation
\begin{equation}
\epsilon( k) =\gamma^2.
\end{equation}
If $\alpha \ne 0$, one has 
\begin{equation}
\epsilon=\frac{-1+\sqrt{1+4\alpha\gamma^2}}{2\alpha}. \label{dispersion}
\end{equation}


The direct use of relation (\ref{dispersion}) into the Schr\"odinger equation (SE) for electron transport is not an easy task. The most straightforward approach consists of writing the SE in the momentum representation but in such a case the analogous formulation of  the open boundary conditions similar to those in the coordinate representation, see for example \cite{Fre}, is still an open problem. In any case, since the parabolic band approximation gives results, which are not very accurate in the semiclassical case \cite{Ro,CaMaRo_book}, it is desirable to include effects due to non-parabolicity in the charge transport. To this aim, we proceed with a perturbative approach. For semiconductors, one has $0<\alpha<1$ (in unit of eV$^{-1}$), so we can expand the dispersion relation as powers of  $\alpha$. Formally we consider $\alpha$ as a {\em small} parameter. Up to the first order, we get
\begin{align}
\epsilon&= \gamma^2-\gamma^4\alpha+o(\alpha^2)=\frac{p^2}{2m^*}-\alpha\frac{p^4}{4(m^*)^2}+o(\alpha^2),
\end{align}
where ${\bf p} = \hbar {\bf k}$ is the crystal momentum and $p$ its modulus.  

By introducing the  momentum operator  expressed in terms of coordinates $P=-i\hbar \nabla$, we have
$$\epsilon\rightarrow -\frac{\hbar^2}{2m^*}\Delta-\frac{\hbar^4}{4(m^*)^2}\alpha\Delta^2+\ldots.$$
Let us  consider the stationary SE 
$$
H \Psi = E \Psi,
$$
    where $\Psi$ is the stationary wave function, $H = \epsilon (P) - q V(X)$ is the Hamiltonian and $E$ is the eigen-energy, with $V({\bf x})$ the electrostatic potential and $q$ elementary (positive) electron charge. 
By inserting the approximation of $\epsilon$ up to first order in $\alpha$, 
we get the following fourth-order stationary Sch\"odinger Equation (SE4) in the coordinate representation:
\begin{equation}
 - \alpha \frac{\hbar^4}{4(m^*)^2}\Delta^2 \Psi  - \frac{\hbar^2}{2m^*}\Delta \Psi - q V({\bf x}) \Psi = E \Psi.\label{SE4}
\end{equation}
In the following, we set $$
a=-\frac{\hbar^4}{4(m^*)^2}\alpha
$$
and assume  the electrostatic potential $V({\bf x})$  as an external (assigned) real field. 

We observe that the approximate energy band
$$
\epsilon( p) = \frac{p^2}{2m^*}-\alpha\frac{p^4}{4(m^*)^2}
$$
vanishes for $p =0$ and for $p =  \pm \displaystyle{\sqrt{\frac{2 m^*}{\alpha}}}$. Therefore, we restrict the momentum to the set  
\begin{equation}
{\cal B} = \left[- \sqrt{\frac{2 m^*}{\alpha}}, \sqrt{\frac{2 m^*}{\alpha}} \right] \label{Brillouin}
\end{equation}
which represents a sort of Brillouin zone.

It is interesting to remark that the fourth-order SE  derived in  \cite{MaDar} following an approach devised by Schr\"odinger  himself differs from (\ref{SE4}) for several aspects. The equation proposed in \cite{MaDar} contains also first and second derivatives of the potential but, more stringently, has solution which are real-valued.  We have only included a more general dispersion relation in the framework of the canonical quantum wave equation, so the  (more precisely pseudo-differential) operator related to the electron energy band acts only on the wave function and no derivatives of the potential arise.

\section{Transparent boundary conditions for the fourth-order Schr\"odinger equation}
\begin{figure}[H]
    \centering
        \includegraphics[width=0.8\textwidth]{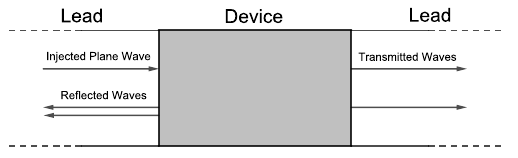} 
        \caption{Schematic representation of an electron device and leads.} \label{fig:device}
\end{figure}

A typical structure of an electron device comprises several parts. A schematization used in several papers \cite{LeKi,AbDeMa,Pi} includes the semiconductor part plus the contacts, which are considered as infinite waveguides, the access zones. In a one-dimensional geometry, which is appropriate, for example, for a resonant tunneling diode, the problem is posed on the real axis which is considered as the union of the half-line $]- \infty, 0 [$ representing the left contact (region $I$), the {\em active} region $[0, L]$ (region $II$) representing the semiconductor device, namely the area where all the relevant physical phenomena occur, and the half-line $]L, + \infty[$ (region $III$) representing the right contact (see Fig. \ref{fig:device}). 
The contacts are metallic, and along them the electrostatic potential is constant. We assume that $V(x)$ is continuous at each interface contact/semiconductor
\begin{equation}
V(x) = \left\{ \begin{array}{lr}
V(0) & x < 0\\
V(L) & x > L
\end{array}
\right. 
\end{equation}
and belonging to $L^{\infty} (\R)$.
In the sequel, we set $V(0) =V_0= 0$ and $V(L)=V_L\geq 0$. 

In principle, given $V(x)$, we have to solve the SE in all $\R$. The idea proposed in \cite{LeKi} and then adopted in several articles \cite{MeJuKo,HeWa,Pi} is to devise appropriate boundary conditions at $x=0$ and $x=L$. In this way, the problem is reduced to a boundary value problem in [0,L]. The resulting SE augmented with the open boundary conditions  is much more feasible for the numerical simulations. 

We split the solution of SE4 as follows:
$$
\Psi(x) = \left\{  \begin{array}{ll}
\Psi_{I}(x) &  x < 0\\
\Psi_{II}(x) &  0\le x \le L\\
\Psi_{III}(x) &  x > L\\
\end{array}
\right.
$$

$\Psi_{I}(x)$ and $\Psi_{III}(x)$ solve in the respective regions a SE with a constant potential.
Electrons can enter the active region from both the contacts: from the left if $p >0$, from the right if $p <0$. They are assumed to be described by a plane wave in the entering waveguide. At the entering boundary they give rise to a reflected wave while at the other boundary to a transmitted one. 

The TBCs have been obtained for the second-order SE (hereafter SE2) for the first time in \cite{LeKi} and several analytical investigations have been performed in \cite{AbDeMa,Ab,Ar,ArNe}. Here we want to generalize such TBCs to the case of SE4.  In region $I$ SE4 admits four plane waves, two of them with positive momentum and the other two with negative momentum. Therefore, a first difference between SE2 and SE4 is that the reflected part of the wave function is a combination of two plane waves instead of one and as a consequence, two reflection coefficients $r_1$ and $r_2$ appear. Similarly, in region $III$ we have two transmitting coefficients $t_1$ and $t_2$. 

In order to get the desired TBCs we impose the continuity of $\Psi$ and its derivatives up to the third order at the edges of the active region. 

%
%
%
%

\begin{remark}
If one imposes the continuity  of $\Psi$ and its derivatives up to the second order at $x = 0$ and $x=L$, the continuity of the third derivatives follows from the equation.
\end{remark}
Indeed, let us consider the one-dimensional fourth-order stationary Schr\"odinger equation
\begin{equation}
a\Psi^{''''}(x) - \frac{\hbar^2}{2m^*}\Psi^{''}(x) - (qV(x)+E)\Psi(x)=0, \quad x \in \R.
\label{SE4d1}
\end{equation}
Let us take $\delta > 0$ and integrate (\ref{SE4d1}) in the interval $[- \delta, \delta]$. One gets
$$
a \left[\Psi^{'''}(x)\right]_{x=-\delta}^{x=\delta} - \frac{\hbar^2}{2m^*} \left[\Psi^{'}(x)\right]_{x=-\delta}^{x=\delta} - \int_{-\delta}^{\delta} (qV(x)+E)\Psi(x) \, d x=0
$$
and, under the considered hypotheses, by taking the limit as $\delta \to 0^+$  the continuity of $\Psi^{'''}$ at $x=0$ follows. Similar results hold at $x=L$.
\begin{remark}
From the point of view of spectral theory for the SE, the natural assumption on the solutions of (\ref{SE4d1}) is that
$$
\Psi \in W^{4,2} (\R)
$$ 
from which it follows the continuity of $\Psi$ and its derivatives up to order three by the Sobolev embedding theorems. 
\end{remark}
\noindent
Here $W^{4,2}(\R)$ represents the Sobolev space
    of the functions $f: \R \to \R$ which admits generalized derivatives up to order $4$ belonging to $L^2(\R)$.
    
\vskip 0.5cm
\noindent  Now we proceed according to the sign of $p$.
\vskip 0.5cm
\noindent
\textbf{Case $p>0$}: \textit{electron waves are injected at $x=0$ and either reflected at $x=0$ or transmitted at $x=L$.} \\

\noindent
We assume the following ansatz:

\begin{equation}
\left\{
\begin{array}{ll}
\displaystyle \Psi_{I}(x)=e^{ik_1x}+r_1e^{-ik_1x}+r_2e^{-ik_2x}, &  x<0\\
\Psi_{III}(x)=t_1e^{ik_3x}+t_2e^{ik_4x}, &  x>L
\end{array}
\right.
\end{equation}

\noindent
where $k_1 = p/\hbar$, with $r_i,t_i\in \C, \ i=1,2$ and $k_i>0, \ i=1,\ldots,4$.

\noindent 
From the continuity of $\Psi$ and its derivatives in $x=0$ up to the third order, one gets

\begin{align}
1 + r_1 + r_2 &= \Psi_{II}(0) \label{cond1} \\
i k_1 - ir_1k_1 - ir_2k_2 &= \Psi^{'}_{II}(0) \label{cond2} \\
-k_1^2 - r_1k_1^2 - r_2k_2^2 &= \Psi^{''}_{II}(0) \label{cond3} \\
-i k_1^3 + ir_1k_1^3 + ir_2k_2^3 &= \Psi^{'''}_{II}(0) \label{cond4}
\end{align}

\noindent 
If  $k_1\neq k_2$ from the conditions (\ref{cond1}), (\ref{cond2}) we find\footnote{For the sake of simplicity, we write $\Psi:=\Psi_{II}$. } the reflection coefficients 

\begin{align}
r_1&=\frac{\Psi'(0)+ik_2\Psi(0)-ik_2-ik_1}{ik_2-ik_1} , \label{r1}\\
r_2&=\frac{-\Psi'(0)-ik_1\Psi(0)+2ik_1}{ik_2-ik_1}. \label{r2}
\end{align}

\noindent 
If we substitute (\ref{r1}), (\ref{r2}) in (\ref{cond3}), (\ref{cond4}) one obtains the boundary conditions at $x=0$
\begin{eqnarray}
i \Psi^{''}(0) - \Psi'(0)(k_1 + k_2) -  i \Psi(0) k_1 k_2  + 2 ik_1 (k_1 + k_2) = 0,
\label{B1}\\
i \Psi^{'''}(0)+i \Psi'(0) (k_1^2 + k_2^2 + k_1 k_2)  - \Psi(0) k_1 k_2 (k_2 + k_1) + 2k_1 k_2 (k_1+k_2) = 0.
\label{B2}
\end{eqnarray}

\noindent  
If $k_1=k_2$ from the conditions (\ref{cond1}), (\ref{cond2})  one has
\begin{equation}
\left\{
\begin{array}{rl}
\displaystyle r_1 + r_2 = \Psi(0) - 1 \\
ik_1(r_1 + r_2) = ik_1 - \Psi'(0)
\end{array}
\right.
\quad \Rightarrow \quad
2ik_1=\Psi'(0)+ik_1\Psi(0)
\end{equation}

\noindent
Similarly from the conditions (\ref{cond3}), (\ref{cond4}) if $k_1=k_2$, one has

\begin{equation}
\left\{
\begin{array}{rl}
\displaystyle k_1^2(-1-r_1-r_2)=\Psi^{''}(0)\\
ik_1^3(-1+r_1+r_2)=\Psi^{'''}(0)
\end{array}
\right.
\quad \Rightarrow \quad
-2ik_1^3=ik_1\Psi^{''}(0)+\Psi^{'''}(0)
\end{equation}
\noindent 
Note that in the case $k_1=k_2$ we have just one reflection coefficient given by  $r_1+r_2$ and the boundary condition at $x =0$ steams as compatibility relation. 

Now we seek the transmission coefficients. From the continuity of $\Psi$ and its derivatives up to third order  in $x=L$ we get 
\begin{align}
t_1e^{ik_3L}+t_2e^{ik_4L}&= \Psi(L), \label{cond11} \\
ik_3t_1e^{ik_3L}+ik_4t_2e^{ik_4L}&= \Psi^{'}(L), \label{cond21} \\
-t_1k_3^2e^{ik_3L}-t_2k_4^2e^{ik_4L} &= \Psi^{''}(L), \label{cond31} \\
-it_1k_3^3e^{ik_3L}-it_2k_4^3e^{ik_4L}  &= \Psi^{'''}(L). \label{cond41}
\end{align}

\noindent 
 If $k_3\neq k_4$ from the conditions (\ref{cond11}), (\ref{cond21}) one has\footnote{For the sake of simplicity, we write $\Psi:=\Psi_{II}$. }
\begin{align}
t_1&=\frac{i k_4\Psi(L)-\Psi'(L)}{ie^{i k_3 L}(k_4-k_3)} , \label{t1}\\
t_2&=\frac{\Psi'(L)-ik_3\Psi(L)}{ie^{i k_4 L}(k_4- k_3)}.\label{t2}
\end{align}

\noindent
If we substitute (\ref{t1}), (\ref{t2}) in (\ref{cond31}), (\ref{cond41}), we find the boundary conditions

\begin{eqnarray}
i\Psi^{''}(L)+\Psi^{'}(L)(k_3 + k_4) - i \Psi(L) k_3 k_4 =0,
\label{B3}\\
\Psi^{'''}(L)+\Psi^{'}(L)(k_3^2 +k_4^2 + k_3 k_4) - i \Psi(L) k_3 k_4 (k_3 + k_4)=0.
\label{B4}
\end{eqnarray}

\noindent
 If $k_3=k_4$ from the conditions (\ref{cond11}), (\ref{cond21}) we get

\begin{equation}
\left\{
\begin{array}{rl}
\displaystyle (t_1+t_2)e^{ik_3L}=\Psi (L)\\
 i(t_1+t_2)k_3e^{ik_3L}=\Psi^{'}(L)
\end{array}
\right.
\quad \Rightarrow \quad
\Psi^{'}(L)=ik_3\Psi(L)
\end{equation}
and similarly from the conditions (\ref{cond31}), (\ref{cond41})
\begin{equation}
\left\{
\begin{array}{rl}
\displaystyle -(t_1+t_2)k_3^2e^{ik_3L}=\Psi^{''}(L)\\
 -i(t_1+t_2)k_3^3e^{ik_3L}=\Psi^{'''}(L)
\end{array}
\right.
\quad \Rightarrow \quad
\Psi^{'''}(L)=ik_3\Psi^{''}(L)
\end{equation}
Note that in the case $k_3=k_4$ as for the reflection coefficients we have just one transmission coefficient given by  $t_1+t_2$ and the boundary condition at $x =L$ steams as compatibility relation.

\vspace{1cm}
Now we pass to evaluate the wave-vectors $k_i$. 
\noindent
In the region $x<0$, after  substituting the ansatz in the Schr\"odinger equation (\ref{SE4d1}) and by using the independence of the function $e^{-ik_ix}$ for different $k_i$'s, one has
\begin{align}
&ak_1^4 + \frac{\hbar^2}{2m^*}k_1^2- (qV(0)+E) = 0, \label{energy1}\\ 
&ak_2^4 + \frac{\hbar^2}{2m^*}k_2^2- (qV(0)+E)  = 0, \label{energy2}
\end{align}

\noindent 
Since $k_1$ is given, from (\ref{energy1}) we find the value of the energy
\begin{equation}
E=\frac{\hbar^2}{2m^*}k_1^2+ak_1^4-qV(0)
\end{equation}
while $k_2$ is the positive solutions of   (\ref{energy2}) distinct from $k_1$ (we recall that $a <0$ and $E > 0$ if $k_1 \in {\cal B}$).

\noindent
On the other hand, using the ansatz in (\ref{SE4d1}), for $x>L$ with arguments similar to the case $x <0$ we get 

%
%

\begin{equation}
k_{3,4} =
\sqrt{-\frac{\hbar^2}{4 m^* a}\pm \frac{1}{2 a}\sqrt{\frac{\hbar^4}{4(m^*)^2}+4a(qV(L)+E)}}.
\end{equation}
Here for $\beta \in \R$, $\sqrt{\beta}$ must be intended as the positive square root if $\beta \ge 0$. If we get complex roots $\pm (\delta + i \sigma)$ then we have to choose the sign that leads to  an evanescent wave in the considered region. Therefore, for $x >L$, we have to take  the sign $+$.
\vskip 0.5cm
\noindent
\textbf{Case $p<0$}: \textit{electron waves are injected at $x=L$ and either reflected at $x=L$ or transmitted at $x=0$.} \\

\noindent
This time we assume the following ansatz
\begin{equation}
\Psi(x) = 
\begin{cases} 
\displaystyle \Psi_{I}(x)= t_1 e^{-ik_3(x-L)} + t_2 e^{-ik_4(x-L)}, & \text{if } x < 0, \\
\Psi_{II}(x), & \text{if } 0 < x < L, \\
\Psi_{III}(x)=e^{ik_1(x-L)} + r_1 e^{-ik_1(x-L)} + r_2 e^{-ik_2(x-L)}, & \text{if } x > L.
\end{cases}
\end{equation}

\noindent
with $r_i,t_i\in \C, \ i=1,2$,and $k_i>0, \ i=1, 2, 3, 4$.

\noindent
With  calculations analogous to the case $p >0$, if $k_1\neq k_2$ and $k_3\neq k_4$ we get the following boundary conditions

\begin{eqnarray}
&& i \Psi^{''}(0) - \Psi^{'}(0)(k_3 + k_4)- i \Psi(0) k_3k_4 = 0,
\label{C1}\\
&& \Psi^{'''}(0)+\Psi^{'}(0)(k_3^2  + k_3 k_4 + k_4^2)+i \Psi(0) k_3 k_4 (k_3 + k_4) = 0,
\label{C2}\\
&& i\Psi^{''}(L) - \Psi^{'}(L)(k_1 + k_2) - i \Psi(L) k_1 k_2  + 2 i k_1 (k_1 + k_2) = 0,
\label{C3}\\
&& i \Psi^{'''}(L)+i \Psi^{'}(L)(k_1^2+k_1 k_2 + k_2^2) - \Psi(L)k_1 k_2 (k_1 +  k_2)+2k_1k_2(k_2+k_1)= 0.\qquad
\label{C4}
\end{eqnarray}

\noindent
If $k_3=k_4$ instead of (\ref{C1}),(\ref{C2}) one has 
\begin{align}
\Psi^{'}(0)&=-ik_3\Psi(0),\\
\Psi^{'''}(0)&=-ik_3\Psi^{''}(0),
\end{align}
\noindent
while if $k_1=k_2$ instead of (\ref{C3}),(\ref{C4}) one has 
\begin{align}
2ik_1&=\Psi^{'}(L)+ik_1\Psi(L),\\
-2ik_1^3&=\Psi^{'''}(L)+ik_1\Psi^{''}(L).
\end{align}

\noindent
Now the energy is given by $E=\displaystyle{\frac{\hbar^2}{2m^*}}k_1^2+ak_1^4-qV(L)$.

As for the case $p >0$, if we get complex roots $\pm (\delta + i \sigma)$ then we have to choose the sign that leads to  an evanescent wave in considered region.  Therefore, for $x >L$, we have to take  the sign $-$.


\section{Well-posedness of the fourth-order Schr\"odinger equation with open boundary conditions}

We want to establish some conditions for the well-posedness of SE4 with the open boundary conditions devised in the previous section. Due to the symmetry of the problem, we will investigate only the the case $p>0$ but the results are valid also when $p <0$. Existence and uniqueness results have been established in \cite{AbDeMa,Ab} for the classical (second-order) Schr\"{o}dinger equation but SE4 is much more complex to tackle, so at the present time we are able to get the same properties only under suitable conditions. These  will be checked numerically in the last section. 

%
%
%
In the sequel, we will denote with $W^{4,1}(0,L)$ the Sobolev space
$$W^{4,1}(0,L)=\{ u\in L^1(0,L):D_{\alpha}u\in L^1(0,L), \forall \alpha, |\alpha| \le 4\}$$
where $D_{\alpha}u$ denotes the generalized derivative of order $\alpha$.

\noindent
\begin{proposition}
Let $V(x) \in L^{\infty}(0,L)$ and be real; let $p\in {\cal B}$ positive be the momentum of an incoming electron from the left waveguide.   If $k_1\neq k_2$ and $k_3\neq k_4$ then  equation (\ref{SE4d1}) with the boundary conditions (\ref{B1}), (\ref{B2}), (\ref{B3}), (\ref{B4}) admits a unique solution belonging to $W^{4,1}(0,L)$ provided that the determinant of the matrix
\begin{equation}
A = \left(
\begin{array}{cccc}
- ik_2k_1 & -(k_1+k_2) &  i & 0\\
-k_1k_2(k_1+k_2) &  i (k_2^2+k_1^2+k_2k_1) & 0 &   i\\
a_{30} & a_{31} & a_{32} & a_{33}\\
a_{40} & a_{41} & a_{42} & a_{43}
\end{array}
\right)
\end{equation}
does not vanishes. Here
\begin{eqnarray*}
\left\{
\begin{array}{l}
a_{3j} = i\varphi_j^{''}(L)+\varphi_j^{'}(L)(k_3+k_4) -ik_3k_4\varphi_j(L),\\[0.3cm]
a_{4j} =  \varphi_j^{'''}(L)+\varphi_j^{'}(L)(k_4^2+k_3^2+k_3k_4) - ik_3k_4(k_4+k_3)\varphi_j(L)
\end{array} \right. \quad j= 0, \ldots, 3 
\end{eqnarray*}
and $\varphi_j (x)$ j= 0, \ldots, 3 are the solutions of the Cauchy problems given by (\ref{SE4d1}) with boundary  conditions
  \vskip 0.3cm
  \begin{equation}
\begin{tabular}{cccc} 
$\varphi_0(0)=1$, & $\varphi_0^{'}(0)=0$, & $\varphi_0^{''}(0)=0$, & $\varphi_0^{'''}(0)=0$;\\
$\varphi_1(0)=0$, & $\varphi_1^{'}(0)=1$, & $\varphi_1^{''}(0)=0$,  & $\varphi_1^{'''}(0)=0$;\\
$\varphi_2(0)=0$, &$\varphi_2^{'}(0)=0$, & $\varphi_2^{''}(0)=1$,  & $\varphi_2^{'''}(0)=0$;\\
$\varphi_3(0)=0$, & $\varphi_3^{'}(0)=0$, & $\varphi_3^{''}(0)=0$, & $\varphi_3^{'''}(0)=1$.
\end{tabular} \label{initial}
\end{equation}
 \label{esistenza}
\end{proposition}
\noindent
\begin{proof}

\noindent
First we observe that under the hypothesis on $V(x)$ the coefficients of the linear equation (\ref{SE4d1}) are in $L^1_{loc}(0,L)$ (the set of locally summable functions in $(0,L)$). Therefore
the Cauchy problem associated to (\ref{SE4d1}) admits a unique solution belonging to $W^{4,1}(0,L)$ \cite{Ha}. The functions $(\varphi_0,\varphi_1,\varphi_3,\varphi_4)$ are a basis of the solutions of equation (\ref{SE4d1}). Therefore, the
general solution of  (\ref{SE4d1}) can be written as a linear combination of $\varphi_i,\ i=0,\ldots,3$, 
$$\Psi (x)=\sum_{j=0}^3 c_j \varphi_j(x)$$
 with $c_0=\Psi (0), \ c_1=\Psi^{'}(0), \ c_2=\Psi^{''}(0), \ c_3=\Psi^{'''}(0)$. Consequently, the boundary conditions  (\ref{B1}), (\ref{B2}), (\ref{B3}), (\ref{B4})  yield the following linear system for the $c_j$'s
\begin{align}
&ic_2-c_1(k_1+k_2)-c_0ik_2k_1=S_1\\
&ic_3+ic_1(k_2^2+k_1^2+k_2k_1)-c_0k_1k_2(k_1+k_2)=S_2\\
&\sum_{j=0}^3\left[ i\varphi_j^{''}(L)+\varphi_j^{'}(L)(k_3+k_4) -ik_3k_4\varphi_j(L)\right]c_j=R_1\\
&\sum_{j=0}^3\left[ \varphi_j^{'''}(L)+\varphi_j^{'}(L)(k_4^2+k_3^2+k_3k_4) - ik_3k_4(k_4+k_3)\varphi_j(L)\right]c_j=R_2
\end{align} \label{boundary}
\noindent with $S_1=-2ik_1(k_1+k_2), \ S_2=-2k_1k_2(k_1+k_2), R_1=0, R_2=0$.
We have the existence of a unique solution if and only if the matrix $A$  is invertible. 
\end{proof}

\begin{remark}
The boundary value problem made of (\ref{SE4d1}) with the boundary conditions (\ref{B1}), (\ref{B2}), (\ref{B3}), (\ref{B4}) constitutes a generalized Sturm-Liouville problem. According to the Fredholm alternative theorem we have either both existence and uniqueness of the solution or the lack of both of them, apart some exceptional cases. 
\end{remark}
\begin{remark}
In the case of a constant electrostatic potential the plane waves $e^{i \frac p \hbar x}$ are solutions of  (\ref{SE4d1}) with the boundary conditions (\ref{B1}), (\ref{B2}), (\ref{B3}), (\ref{B4}), as it is possible to verify  by a simple direct calculation. 
\end{remark}
\noindent
Analogous results can be obtained if $k_1=k_2$ and/or $k_3=k_4$. 

If $k_1 = k_2$ and $k_3 \ne k_4$ the boundary conditions give rise to the following linear system:
\begin{align}
&c_1+ik_1c_0=S_1 \label{BCeq1}\\
&c_3+ic_1k_1c_2=S_2 \label{BCeq2}\\
&\sum_{j=0}^3\left[ i\varphi_j^{''}(L)+\varphi_j^{'}(L)(k_3+k_4) -ik_3k_4\varphi_j(L)\right]c_j=R_1\\
&\sum_{j=0}^3\left[ \varphi_j^{'''}(L)+\varphi_j^{'}(L)(k_4^2+k_3^2+k_3k_4)-ik_3k_4(k_4+k_3)\varphi_j(L)\right]c_j=R_2
\end{align}
where
$S_1=2ik_1,\ S_2=-2ik_1^3$.

If  $k_1 \ne k_2$ and $k_3 = k_4$ the boundary conditions give rise to the following linear system:
\begin{align}
&ic_2-c_1(k_1+k_2)-c_0ik_2k_1=S_1\\
&ic_3+ic_1(k_2^2+k_1^2+k_2k_1)-c_0k_1k_2(k_1+k_2)=S_2\\
&\sum_{j=0}^3\left[ \varphi_j^{'}(L)-ik_3\varphi_j(L)\right]c_j=R_1 \label{BCeq7}\\
&\sum_{j=0}^3\left[ \varphi_j^{'''}(L)-ik_3\varphi_j^{''}(L)\right]c_j=R_2. \label{BCeq8}
\end{align}
where
with $S_1=-2ik_1(k_1+k_2), \ S_2=-2k_1k_2(k_1+k_2), R_1=0, R_2=0$.

In the case $k_1 = k_2$ and $k_3 \ne k_4$, the boundary conditions lead to equations (\ref{BCeq1}), (\ref{BCeq2}), (\ref{BCeq7}), (\ref{BCeq8}).

\subsection{Example: step function potential}

As an example of application of Proposition \ref{esistenza} we consider a  potential having the following shape
$$
V(x) = \left\{
\begin{array}{ll}
0 & \mbox{if } x \in [0, \frac{L}{2}[\\[0.2cm]
V_L  & \mbox{if } x \in [\frac{L}{2}, L]
\end{array}
\right. 
$$
 Observe that a step potential naturally appears in hetero-junctions made by putting together two semiconductor materials possessing different work functions.  

\noindent
For SE2 we seek the fundamental solutions $\varphi_j,\ j=0,1$ having the form

\begin{equation}
\varphi_{j}(x)=\left\{
\begin{array}{ll}
 d_{j,1} e^{i\frac{p}{\hbar}x}+d_{j,2} e^{-i\frac{p}{\hbar}x} & \mbox{if }  x\in[0,\frac{L}{2}]\\
 m_{j,1} e^{i\frac{p_+}{\hbar}x}+m_{j,2} e^{-i\frac{p_+}{\hbar}x} & \mbox{in } x\in[\frac{L}{2},L]
\end{array}
\right.
\end{equation}
with coefficients $d_{j,1},d_{j,2}, m_{j,1}, m_{j,2}$ to be determined. 
\noindent
If we impose $\varphi^{(i)}_{j}(0)=\delta_{ij},\ i=0,1$ we get $d_{j,i} ,\ i, j =0,1$ whereas, if we impose $\varphi^{(i)}_{j}((\frac{L}{2})^-)=\varphi^{i}_{j}((\frac{L}{2})^+),\ i, j=0,1$, we get $m_{j,i},\ i, j=0,1$.


Similarly, for SE4 we look for fundamental solutions $\varphi_j,\ j=0,1,2,3$ of the type 

\begin{equation}
\varphi_{j}(x)=\left\{
\begin{array}{ll}
d_{j,1} e^{ik_1x}+d_{j,2} e^{ik_2x}+d_{j,3} e^{-ik_1x}+d_{j,4} e^{-ik_2x} & \mbox{if } x\in[0,\frac{L}{2}]\\
 m_{j,1} e^{ik_3x}+m_{j,2} e^{ik_4x}+m_{j,3} e^{-ik_3x}+m_{j,4} e{-ik_4x} & \mbox{if } x\in[\frac{L}{2},L]
\end{array}
\right.
\end{equation}
with coefficient $d_{i,j}$ and $m_{i,j}$,  $i, j = 0,1,2,3$, to be determined. 
\noindent
If we impose $\varphi^{(i)}_{j}(0)=\delta_{ij},\ i,j =0,1,2,3$, we get the following system 
\[
A {\bf d_j} = {\bf b_j}
\]
where
\[
A =
\begin{bmatrix}
1 & 1 & 1 & 1 \\
ik_1 & ik_2 & -ik_1 & -ik_2 \\
-k_1^2 & -k_2^2 & -k_1^2 & -k_2^2 \\
-ik_1^3 & -ik_2^3 & ik_1^3 & ik_2^3
\end{bmatrix},
\quad
{\bf d_j} =
\begin{bmatrix}
d_{j,1} \\
d_{j,2} \\
d_{j,3} \\
d_{j,4}
\end{bmatrix},
\quad
{\bf b_j} =
\begin{bmatrix}
0 \\
\vdots \\
1 \\
\vdots \\
0
\end{bmatrix}
 \leftarrow j \]

\begin{figure}[H]
    \centering
    \subfloat[Case $V_0= 0$ V, $V_L= -0.1$ V.]{
        \includegraphics[width=0.7\textwidth]{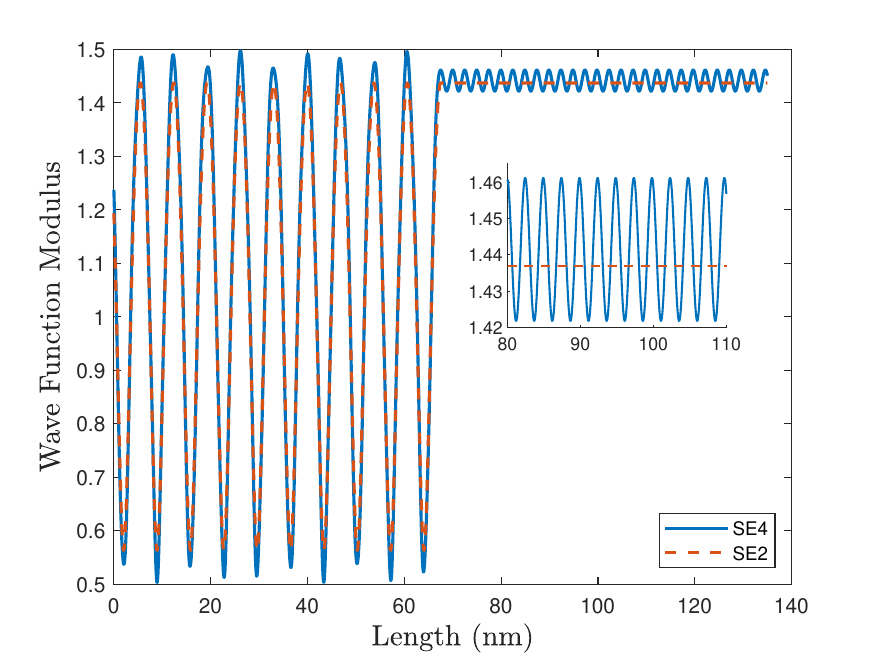}
    }
    \hspace{0.05\textwidth} 
    \subfloat[Case $V_0= 0$ V,\ $V_L= 0.3$ V.]{
        \includegraphics[width=0.7\textwidth]{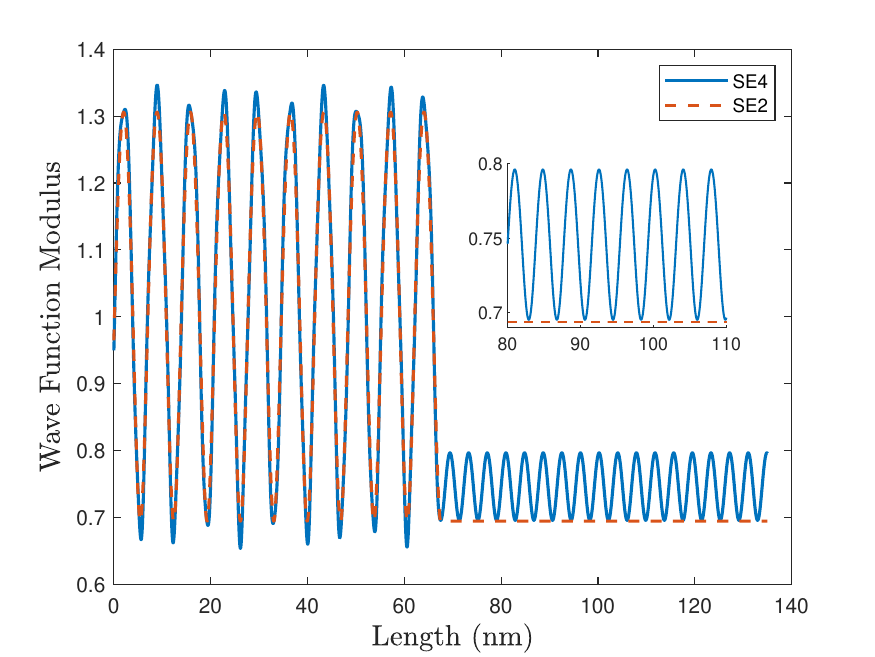}
    }
    \caption{Comparison between the analytical solutions of SE2 and SE4 in the presence of a stepwise potential setting $k_1= 0.4558$ nm$^{-1}$, $\alpha= 0.242$ eV$^{-1}$, $L=135$ nm.}
    \label{StepWise}
\end{figure}

Then if we impose $\varphi^{(i)}_{j}((\frac{L}{2})^-)=\varphi^{(i)}_{j}((\frac{L}{2})^+),\ i, j = 0,1,2,3$, we get also the coefficients $m_{j,i},\ i=1,2,3,4$,


 In Fig. \ref{StepWise}, the modulus of the wave-function in the case $V_L=-0.1$ V (a) and and  $V_L=0.3$ V (b) with incoming wave vector $k_1=$ 0.4558 nm$^{-1}$ is shown by comparing the numerical solution of SE4 with that of  SE2. We set $\alpha =$ 0.242 eV$^{-1}$ and $m^* = 0.067 m_e$, which are appropriate for electrons of the $L$-valley in GaAs,  and  $L=135$ nm.
 Here and in the following section we consider typical times of the order of one femtosecond (fs) and energy of the order of $k_B T$, with $T$ absolute temperature. Assuming a room temperature (300 K) we get for the Planck constant the value $\hbar = 0.6582$ eV fs.
 
 The most evident difference is the fact that in the region 
 $[\frac{L}{2}, L]$ SE2 predicts a plane wave, which has a constant modulus, while the solution of SE4 is the superposition of two plane waves which produces interference, as well evident from Fig. \ref{StepWise}. The difference is more marked in the case $V_L = $ 0.3 V. Instead, the solutions to SE2 and SE4 are in good agreement in the interval $ [0, \frac{L}{2}]$, although there are some slight discrepancies in the extrema.
 
 Even though simple, this example clearly reveals that the non-parabolic corrections to the dispersion relation can produce interesting effects such as the observed interference. The passage through  the device transforms an incoming electron  in a coherent state described by a simple wave to an outgoing scattering state which is the superposition of two plane waves.

\section{Expression of the probability current}
The evolution equation for the square modulus of the wavefunction can be written in a divergence form with a flux which represents for unit charge the  current of the single electron.  
We want to derive the expression of the current for  the SE4 in the spirit of the Ehrenfest's theorem. 

The non stationary version of (\ref{SE4}) in general dimension reads 
\begin{equation}
i\hbar \frac{\partial \Psi}{\partial t}({\bf x},t) =  - \alpha \frac{\hbar^4}{4(m^*)^2}\Delta^2 \Psi  ({\bf x},t) -\frac{\hbar^2}{2m^*} \Delta \Psi({\bf x},t)-qV({\bf x})\Psi({\bf x},t).          \label{TSE4}
\end{equation} 
\begin{proposition}
The evolution of  $|\Psi  ({\bf x},t)|^2$ satisfies the equation in divergence form
\begin{equation}
\frac{\partial }{\partial t}  |\Psi  ({\bf x},t)|^2 + \nabla \cdot {\bf J} = 0\ ,
\end{equation}
where ${\bf J}$ represents the single electron probability current density for unit charge
\begin{equation}
{\bf J} =  \mathfrak{Im} \left(  \frac{\hbar}{m^*}  \overline{\Psi }\nabla \Psi   + \frac{\alpha \hbar^3}{2 m^*}\left( \overline{\Psi } \nabla \Delta^2 \Psi  - \nabla  \overline{\Psi} \Delta^2 \Psi \right)\right).
\label{J}
\end{equation}
\end{proposition}
{\em Proof}. Multiplying (\ref{TSE4}) by $\overline{\Psi }$ one has
\begin{equation}
i\hbar \overline{\Psi } \frac{\partial \Psi}{\partial t}({\bf x},t) =  - \alpha \frac{\hbar^4}{4(m^*)^2} \overline{\Psi }\Delta^2 \Psi  ({\bf x},t) -\frac{\hbar^2}{2m^*} \overline{\Psi } \Delta \Psi({\bf x},t)-qV({\bf x}) |\Psi  ({\bf x},t)|^2.          \label{TSE4_1}
\end{equation} 
By taking the complex conjugate of (\ref{TSE4_1}) we have
\begin{equation}
- i\hbar \Psi \frac{\partial \overline{\Psi } }{\partial t}({\bf x},t) =  - \alpha \frac{\hbar^4}{4(m^*)^2} {\Psi }\Delta^2 \overline{\Psi }   ({\bf x},t) -\frac{\hbar^2}{2m^*} {\Psi } \Delta 
\overline{\Psi } ({\bf x},t)-qV({\bf x}) |\Psi  ({\bf x},t)|^2         \label{TSE4_2}
\end{equation} 
and by subtracting one gets
\begin{eqnarray}
i\hbar\frac{\partial |\Psi|^2}{\partial t}({\bf x},t) + \alpha \frac{\hbar^4}{4(m^*)^2} \left(\overline{\Psi }\Delta^2 \Psi  ({\bf x},t)
 - {\Psi }\Delta^2 \overline{\Psi }   ({\bf x},t)\right)  + \frac{\hbar^2}{2m^*} \left(\overline{\Psi } \Delta \Psi({\bf x},t) - {\Psi } \Delta 
\overline{\Psi } ({\bf x},t)  \right) = 0.
\end{eqnarray}
We observe that
\begin{eqnarray*}
& &\overline{\Psi }\Delta \Psi  ({\bf x},t)
 - {\Psi }\Delta \overline{\Psi }   ({\bf x},t) = \nabla \cdot \left(\overline{\Psi }\nabla \Psi  ({\bf x},t) - \Psi \nabla \overline{\Psi}  ({\bf x},t) \right)\\
& &\overline{\Psi }\Delta^2 \Psi  ({\bf x},t)
 - {\Psi }\Delta^2 \overline{\Psi }   ({\bf x},t) =\\
 & & \nabla \cdot \left(\overline{\Psi } \nabla  \left(\Delta \Psi  ({\bf x},t) \right)  - \Psi \nabla \cdot \left(\Delta  \overline{\Psi}  ({\bf x},t)\right) \right) 
 - \nabla  \overline{\Psi} \cdot \nabla  \left(\Delta \Psi  ({\bf x},t) \right) +  \nabla  \Psi \cdot \nabla  \left(\Delta \overline{\Psi}  ({\bf x},t) \right).
\end{eqnarray*}
Since 
\begin{eqnarray*}
\nabla  \overline{\Psi} \cdot \nabla  \left(\Delta \Psi  ({\bf x},t) \right) = \nabla  \cdot \left( \nabla  \overline{\Psi} \cdot  \Delta \Psi  ({\bf x},t) \right) - \Delta  \overline{\Psi} \Delta \Psi,\\
\nabla  \Psi \cdot \nabla  \left(\Delta  \overline{\Psi}  ({\bf x},t) \right) = \nabla  \cdot  \left( \nabla  \Psi \cdot  \Delta  \overline{\Psi}   ({\bf x},t) \right) - \Delta  \overline{\Psi} \Delta \Psi,
\end{eqnarray*}
the proof is complete.  \hfill $\Box$

In the one-dimensional case the current reads
\begin{equation}
J = \mathfrak{Im} \left(  \frac{\hbar}{m^*}  \overline{\Psi }\Psi^{'}   + \frac{\alpha \hbar^3}{2 (m^*)^2}\left( \overline{\Psi }  \Psi^{'''}  -  \overline{\Psi}^{'}  \Psi^{''} \right)\right).
\label{J1}
\end{equation}
In the limit $\alpha \to 0^+$ the classical expression of ${\bf J}$ is recovered.

Observe that smoothness of the solutions is guaranteed up to third order derivatives, so $J$ is well-defined. 

\begin{figure}[H]
    \centering
        \includegraphics[width=0.7\textwidth]{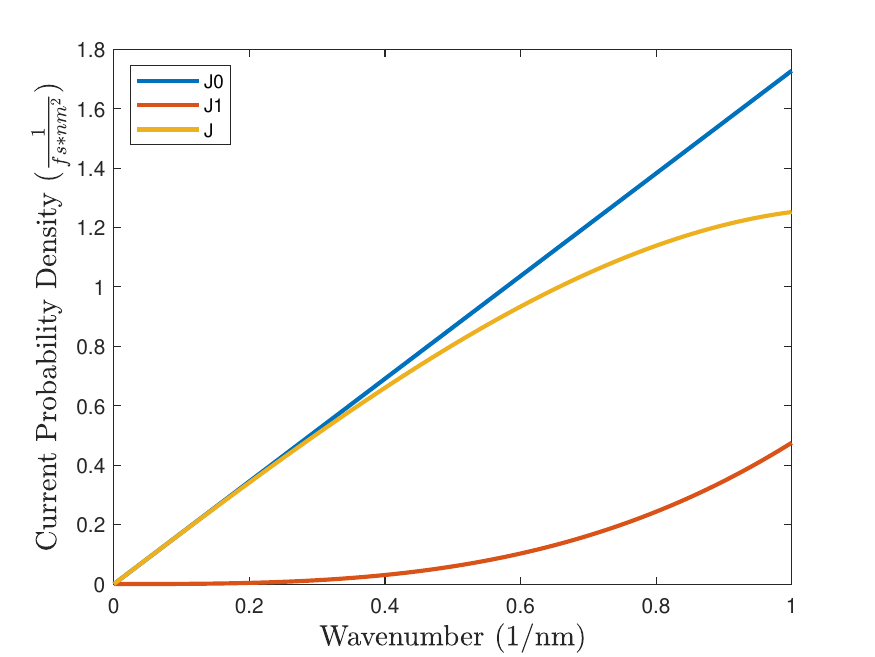} 
        \caption{Current of the plane wave  versus the wave vector $k$ when $\alpha=$ 0.242 eV$^{-1}$} \label{fig:current}
\end{figure}

As example, let us consider a plane wave $\Psi=e^{ikx}$ with $k=\frac{p}{\hbar}$. In such a case
 $J$ can be explicitly written as a sum of two components $J_0,J_1$ with $J_0=\frac{\hbar k}{m}$ and $J_1=-\alpha \frac{\hbar^3k^3}{m^{*2}}$.
In Fig. \ref{fig:current}, the values of $J$ versus the wave-vector are plotted setting $\alpha = 0.242$ eV$^{-1}$.  We see that the current is lower than that one has in the effective mass approximation, even if the difference is significant for values of $k$ greater than about 0.4 nm$^{-1}$.

We remark that a generalization of the expression of the current which takes account the transport of spin with a Pauli formalism can be found in \cite{BoDroFiWe}.


\section{Numerical simulations}

Apart the theoretical interest, Proposition \ref{esistenza} furnishes a very efficient way to solve numerically eq. (\ref{SE4}). We solve the latter with initial conditions (\ref{initial})  by using a high order 
numerical method for ODEs getting the basic functions $\varphi_r(x)$, $r =0,1,2,3$, along with their derivatives up to order three. Specifically, the MATLAB \cite{a} implementation of the Verner's "most robust" Runge-Kutta 9(8) pair with an 8th-order continuous extension \cite{b} has been adopted.
By inserting the values of the functions $\varphi_r(x)$ in the boundary conditions (\ref{boundary}) one obtains the coefficients $c_r$,  $r=0,1,2,3$, and therefore the solution of eq. (\ref{SE4}). As byproduct, since also the derivatives up to order three are obtained because one has to rewrite the Schr\"{o}dinger equation as a system of first order ODEs, it is straightforward to evaluate the current (\ref{J}).

A crucial point is that the potential $V(x)$ may have discontinuities with a consequent loss of regularity. In the cases we are going to investigate, $V(x)$ is piecewise smooth, so we solve eq. (\ref{SE4}) with initial conditions (\ref{initial}) in each interval of regularity of the potential matching the initial data in each subinterval.  More in detail, let us suppose that
there exist disjoint sub-intervals 
$$
I_0 = [0, a_1), I_1 = (a_1,a_2), \ldots, I_r = (a_r,L]
$$ 
which form a partition of $[0, L]$, and let us suppose that $V_{/ I_j}(x)$ is smooth.

First we solve eq. (\ref{SE4}) in the interval $\overline{I_0 }$ with initial conditions (\ref{initial}); then we solve eq. (\ref{SE4}) in the interval $\overline{I_1}$ with initial conditions given by the numerical solution in the previous interval evaluated in $x=a_1$; then we solve eq. (\ref{SE4}) in the interval $\overline{I_2}$ with initial conditions given by the numerical solution in the previous interval evaluated in $x=a_2$, and so on.

This scheme reveals to be very accurate and efficient, as it will be shown in the subsequent examples. In all the simulations of this section we set 
$\alpha=0.242$ eV$^{-1}$, $m^* = 0.067 m_e$ and $\hbar = 0.6582$ eV fs.

\subsection{Single potential barrier}

\begin{figure}[H]
\centering
        \includegraphics[width=0.7\textwidth]{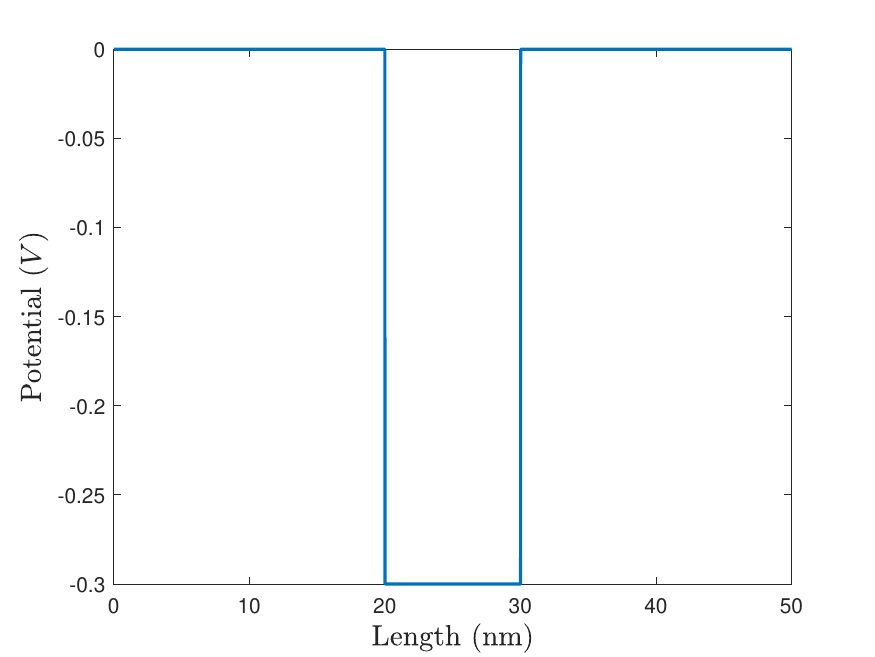}
        \caption{Single barrier} \label{fig:barrier}
\end{figure}

In this case the potential is given by (see Fig. \ref{fig:barrier})
$$
V(x) = \left\{
\begin{array}{ll}
    0 & \mbox{if } x \in [0, a_1[ \cup ]a_2, L]\\[0.2cm]
V_b  & \mbox{if } x \in [a_1, a_2]
\end{array}
\right. 
$$
where $a_1 =  20$ nm, $a_2 = 30$ nm, $L = 50$ nm, $V_b =  -0.3$ V.

\begin{figure}[H]
    \centering
    \subfloat[Case $k_1= 0.7264$ nm$^{-1}$]{
        \includegraphics[width=0.7\textwidth]{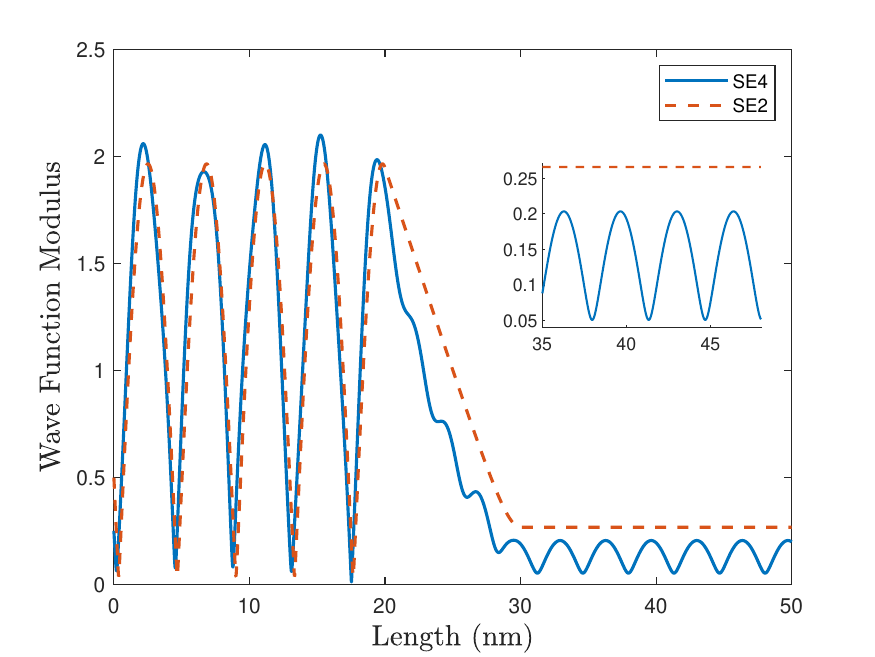}
    }
    \hspace{0.05\textwidth} 
    \subfloat[Case $k_1= 1.064$ nm$^{-1}$]{
        \includegraphics[width=0.7\textwidth]{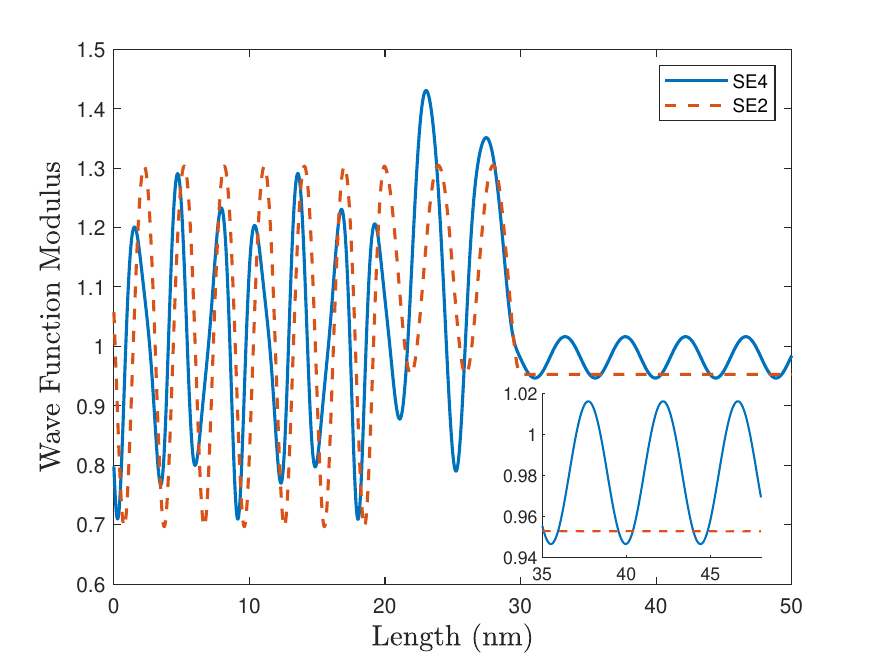}
    }
    \caption{Comparison between the solution of SE2 and SE4 in the case of a single barrier with $V_b= -0.3$ V and $L=50$ nm.}
    \label{fig:sol_barrier}
\end{figure}

The results are illustrated in Fig.s \ref{fig:sol_barrier} for an incoming wave-vector $k_1 = $ 0.7264 nm$^{-1}$ (a) and $k_1=$ 1.064 nm$^{-1}$ (b). We
find the same difference as for the step potential: the solutions of SE4 and SE2 differ mainly beyond the barrier with the effect of modulation which is missing in SE2.  At higher $k_1$ the peaks in the region $[0,a_2]$ are a bit more pronounced.

\subsection{Double potential barrier}

\begin{figure}[H]
   \centering
        \includegraphics[width=0.7\textwidth]{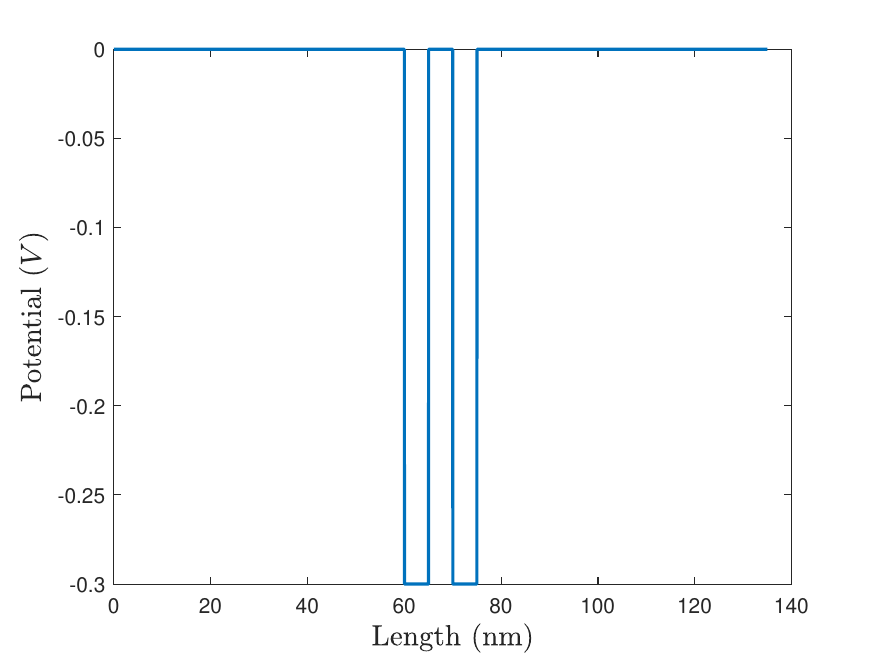}
        \caption{Double barrier.}
    \label{fig:double_barrier}
\end{figure}

The potential is described by the function (see Fig. \ref{fig:double_barrier})
$$
V(x) = \left\{
\begin{array}{ll}
    0 & \mbox{if } x \in [0, a_1[ \cup ]a_2, a_3]\cup [a_4, L]\\[0.2cm]
V_b  & \mbox{if } x \in [a_1, a_2] \cup [a_3, a_4]
\end{array}
\right. 
$$
where $a_1 = $ 60 nm, $a_2 =$ 65 nm, $a_3 = $ 70 nm, $a_4 =$ 75 nm, $L =$ 135 nm, $V_b = - $0.3 V.

The results are illustrated in Fig.s \ref{fig:double_barrier} for an incoming wave-vector $k_1= $ 0.2846 nm$^{-1}$ (a). We
find similar differences as for the single barrier.
  
\begin{figure}[H]
    \centering
    \subfloat[Comparison between the solution of SE2 and SE4 in the case $V_b=-0.3$ V, $k_1=0.2846$ nm$^{-1}$, $\alpha= 0.242$ eV$^{-1}$, $L=135$ nm.]{
        \includegraphics[width=0.7\textwidth]{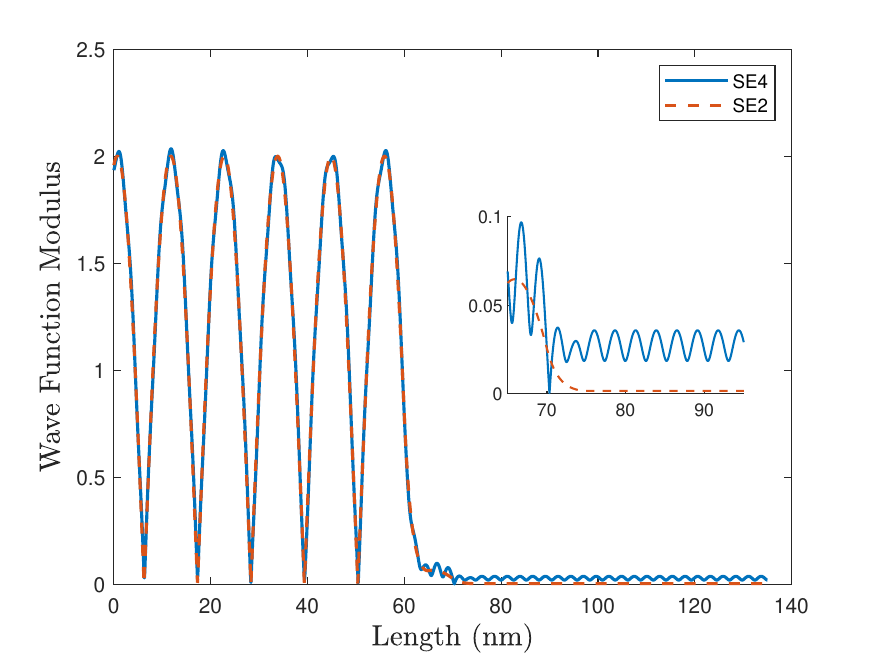}
        \label{fig:sol_barrier1}
   }     
   
    \centering
    \subfloat[Double Barrier  $V_0=0$ V, $V_L=0$, $V_b=-0.3$ V, $k_1= 1.1386$ nm$^{-1}$, $\alpha=0.242$ eV$^{-1}$, $L=135$ nm.]{
        \includegraphics[width=0.7\textwidth]{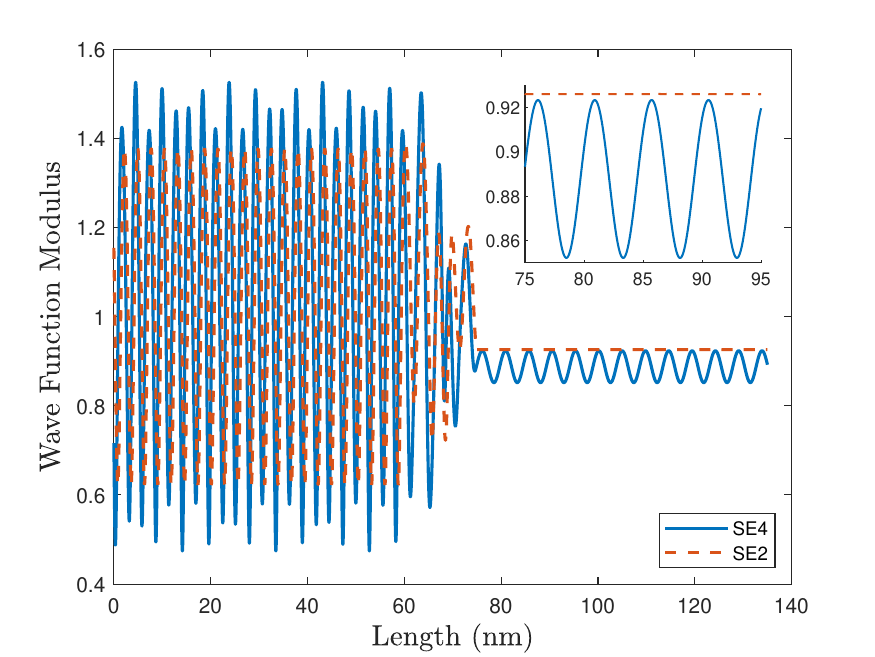} 
    }
        \caption{Comparison between the solution of SE2 and SE4 in the case of a double barrier.} 
         \end{figure}

\subsection{A resonant tunneling diode type potential }

\begin{figure}[H]
    \centering
{
        \includegraphics[width=0.7\textwidth]{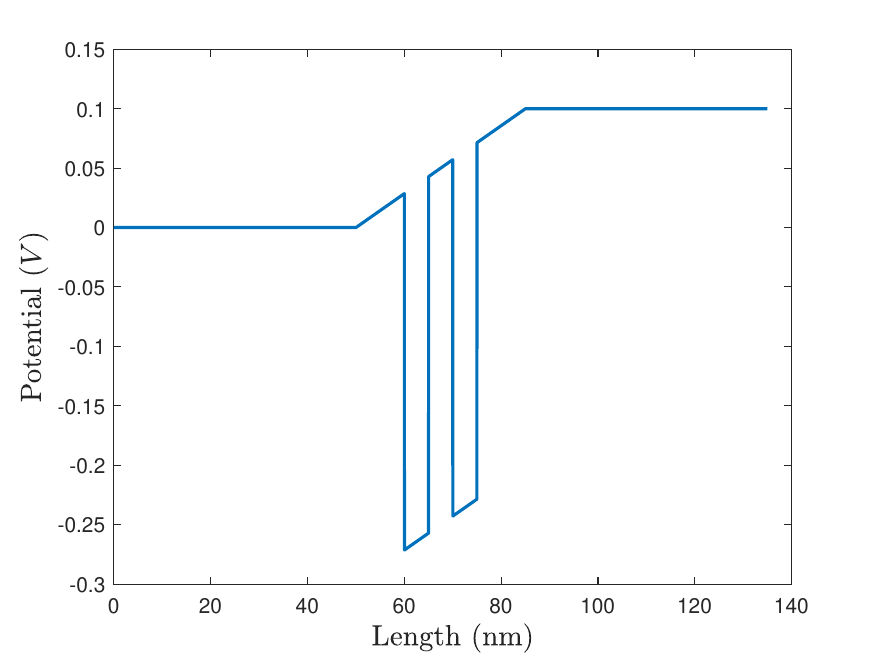}
       }
 \caption{RTD type potential in the case $V_0=0$ V, $V_L= 0.1$ V, $V_b = -0.3$ V,  $L=135$ nm.}
    \label{fig:rampa}
\end{figure}

\begin{figure}[H]
    \centering
{
        \includegraphics[width=0.7\textwidth]{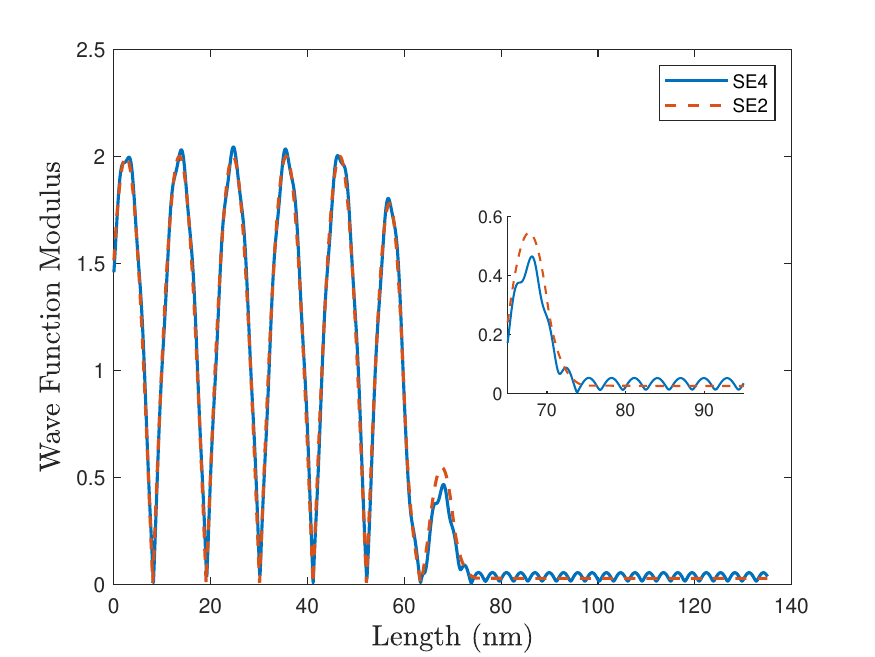}    
  }      
    \caption{Comparison between the solution of SE4 and SE2  for the  resonant tunneling diode type of Fig. 
    \ref{fig:rampa} in the case $k_1=0.2846$ nm$^{-1}$, $\alpha=0.242$ eV$^{-1}$. }
    \label{fig:RTD}
\end{figure}

\begin{figure}[H]
    \centering
\subfloat[RTD type potential:  $V_0=0$ V, $V_L= 0.1$ V, $V_b = -0.3$ V, $k_1=0.2846$ nm$^{-1}$, $\alpha=0.242$ eV$^{-1}$, $L=135$ nm.]{
        \includegraphics[width=0.7\textwidth]{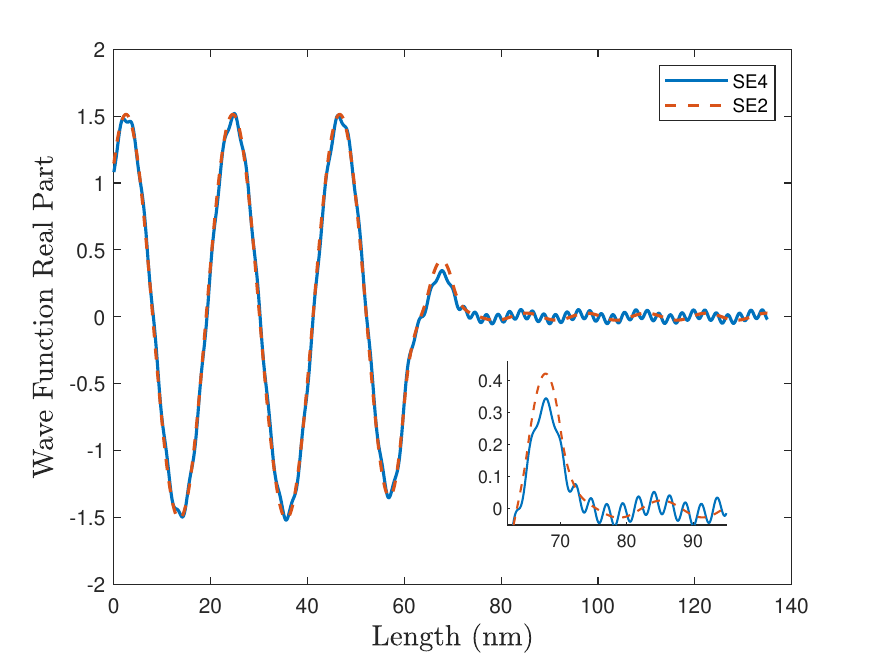}
       }
    \hspace{0.05\textwidth} 
 \subfloat[RTD type potential: $V_0=0$ V, $V_L= 0.1$ V, $V_b = -0.3$ V, $k_1=0.2846$ nm$^{-1}$, $\alpha=0.242$ eV$^{-1}$, $L=135$ nm.]{
        \includegraphics[width=0.7\textwidth]{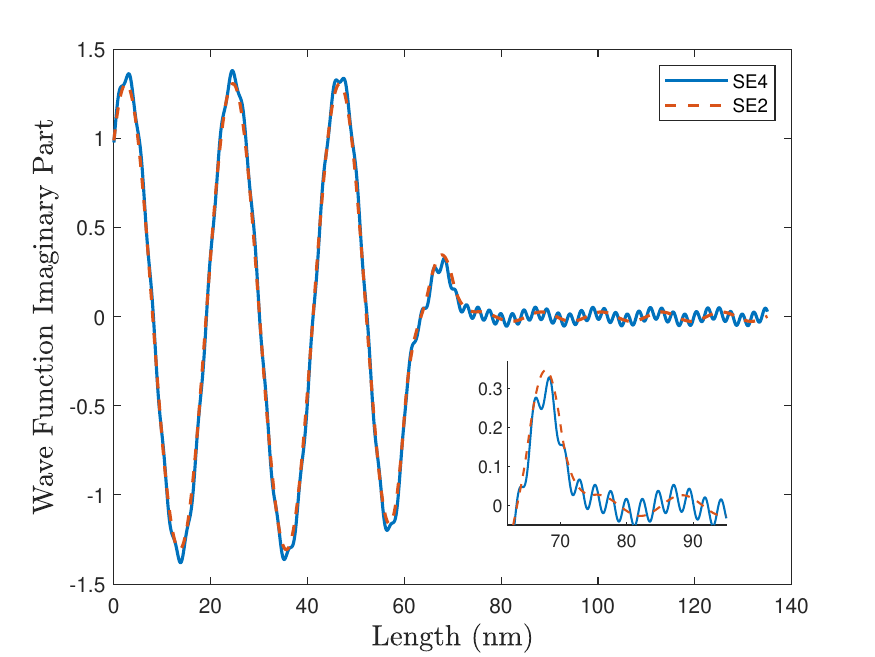}    
    }
    \caption{Comparison between the real and imaginary parts of solution of SE4 and SE2.}
    \label{fig:RTD_Re_Im}
\end{figure}

\begin{figure}[H]
    \centering
{
        \includegraphics[width=0.7\textwidth]{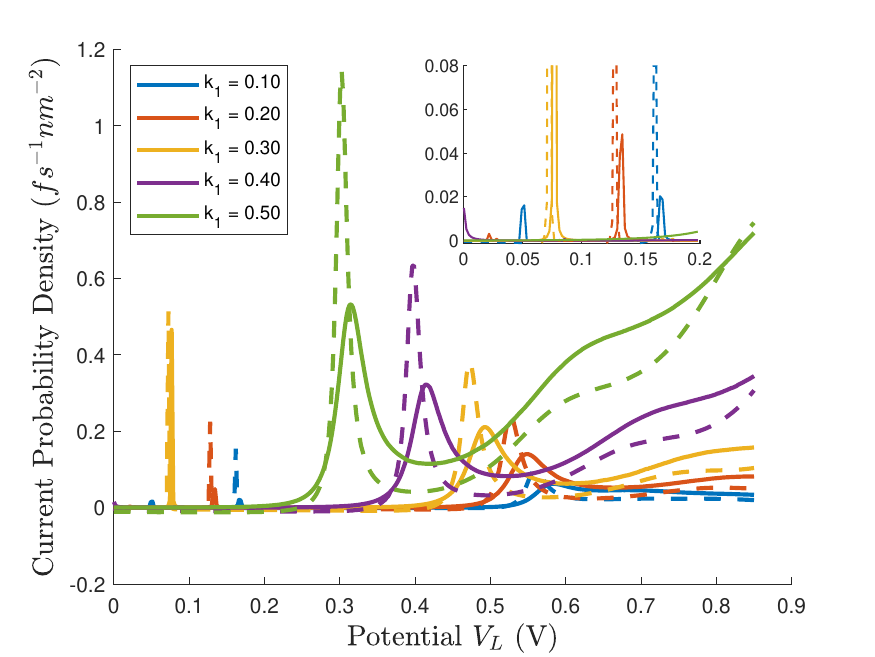}
       }
   
    \caption{Characteristic curves (probability current density per unit charge versus $V_L$) for SE4 (dashed line) and SE2 (continuous  line).}
    \label{fig:IV}
\end{figure}

An expression of $V(x)$ which resembles the typical behavior in a resonant tunneling diode (RTD), at least qualitatively, is that proposed in \cite{HeWa}  (see Fig. \ref{fig:rampa})
$$
V(x) = \left\{
\begin{array}{ll}
    0 & \mbox{if } x \in [0, a_1[ \\[0.2cm]
   (x - a_1) \frac{V_L}{a_6 - a_1} &  \mbox{if } x \in [a_1, a_2]\cup \in [a_3, a_4]\cup [a_5, a_6] \\[0.2cm]
  (x - a_1) \frac{V_L}{a_6 - a_1} + V_b & \mbox{if } x \in ]a_2, a_3[ \cup ]a_4, a_5[\\[0.2cm]
V_L & \mbox{if } x \in ]a_6,L]
\end{array}
\right. 
$$
which is the superposition of a double barrier and a linear potential.
In our simulations we set $a_1 = $ 50 nm, $a_2 =$ 60 nm, $a_3 = $ 65 nm, $a_4 =$ 70 nm, $a_5 = $ 75 nm, $a_6 =$ 85 nm, $L =$ 135 nm,  $V_b =-0.3$ V. The wave-vector of the entering wave is taken as $k_1=$ 0.2846 nm$^{-1}$. Again the main difference between the solutions of SE2 and SE4 is in the last part of the device where an interference effect is present if the SE4 is adopted in the description as shown in Fig.s \ref{fig:RTD}, \ref{fig:RTD_Re_Im}.  Numerically we have got a good conservation of $J$ confirming the accuracy of the numerical scheme. In Fig.s \ref{fig:IV} the current versus $V_L$ is depicted. It is evident we have resonance for some values of the wave vector and potential $V_L$. The qualitative behavior is the same for the SE2 and SE4 and, notably, one observes resonances. When there is resonance for  both SE2 and SE4, the  wave vectors  of resonance are close and, of course, they are observed for low bias voltages $V_L$. As $V_L$ increases we have an effect of negative differential conductivity; namely, there is a local maximum after which the current decreases in a certain range of $V_L$ before increasing again for large $V_L$. Apparently an exception is given by the solution for $k_1 = 0.1$ nm$^{-1}$ because after the local maximum the current is monotonically decreasing at least for the considered values of $V_L$. 

\section*{Conclusions and acknowledgments}

A dispersion relation beyond the effective mass approximation has been included in the Schr\"{o}dinger equation for a single electron. This leads to a hierarchy of Schr\"{o}dinger equations of increasing order. A detailed analysis has been performed in the case of the fourth-order Schr\"{o}dinger equation for describing charge transport in a semiconductor device. An explicit relation of the probability current density has been obtained and appropriate transparent boundary conditions have been devised to reduce the problem to a generalized Sturm-Liouville problem in a finite spatial domain. Conditions for the well-posedness of the resulting boundary value problem have been formulated. The shown examples of solutions  highlight that the fourth-order Schr\"{o}dinger equation gives rise to interesting effects of resonance which are missing from SE2.
As future application, SE4 will be employed for the simulation of a resonant tunneling diode. 

The authors acknowledge the support from INdAM (GNFM) and from MUR progetto PRIN
{\it Transport phonema in low dimensional structures: models, simulations and theoretical aspects} 
CUP E53D23005900006.

\clearpage \noindent
%


\end{document}